\documentclass[aps, prd, twocolumn, superscriptaddress, showpacs, preprintnumbers, amsmath, amssymb, tightenlines]{revtex4}

\usepackage{graphicx} % Include figure files
\usepackage{dcolumn}  % Align table columns on decimal point
\usepackage{subfloat}
\usepackage[caption=false]{subfig}
\usepackage{hyperref}
\usepackage{bm}
\usepackage{color}
\graphicspath{{ps}}

\renewcommand{\arraystretch}{1.1}

\begin{document}

%\preprint{\vbox{ \hbox{   }
%			 \hbox{Belle DRAFT {\it 2013-03-12}}
%                        \hbox{Intended for {\it PRD(RC)}}
%                        \hbox{Author: O.Lutz(Brovchenko)}
%                        \hbox{Committee: Youngjoon Kwon (chair),}
%                        \hbox{Kai-Feng Chen, Shoji Uno, }
  		              % \hbox{hep-ex nnnn}
%}}

\title{ \quad\\[1.0cm] Search for \bm{$B\rightarrow h^{(*)}\nu\bar{\nu}$} with the full Belle $\Upsilon(4S)$ data sample}

%%%% >>>>> insert the authorlist here. BEFORE the abstract !!!!! <<<<<
%%%% >>>>> from the authorship confirmation web page
%%% Name the file author.tex and use \input{author} to insert into your latex file.
%%% Paper:    B -> h nu nubar
%%% Journal:  Physical Review D (Rapid Communication)
%%% Contacts: O.Lutz (oksana.brovchenko@kit.edu)
%%% Non-responding authors or those who said NO are commented out.
%%% ====================================================================
%%% Click the RELOAD button on your web browser to see the updated file.
%%% ====================================================================
%%% Use \input{author} to insert this material into your latex file.
%%%%% Force institutions to appear in alphabetical order when typeset.
\noaffiliation
\affiliation{University of the Basque Country UPV/EHU, 48080 Bilbao}
\affiliation{University of Bonn, 53115 Bonn}
\affiliation{Budker Institute of Nuclear Physics SB RAS and Novosibirsk State University, Novosibirsk 630090}
\affiliation{Faculty of Mathematics and Physics, Charles University, 121 16 Prague}
%%%\affiliation{Chiba University, Chiba 263-8522}
\affiliation{University of Cincinnati, Cincinnati, Ohio 45221}
%%%\affiliation{Department of Physics, Fu Jen Catholic University, Taipei 24205}
\affiliation{Justus-Liebig-Universit\"at Gie\ss{}en, 35392 Gie\ss{}en}
\affiliation{Gifu University, Gifu 501-1193}
%%%\affiliation{II. Physikalisches Institut, Georg-August-Universit\"at G\"ottingen, 37073 G\"ottingen}
%%%\affiliation{The Graduate University for Advanced Studies, Hayama 240-0193}
%%%\affiliation{Gyeongsang National University, Chinju 660-701}
\affiliation{Hanyang University, Seoul 133-791}
\affiliation{University of Hawaii, Honolulu, Hawaii 96822}
\affiliation{High Energy Accelerator Research Organization (KEK), Tsukuba 305-0801}
%%%\affiliation{Hiroshima Institute of Technology, Hiroshima 731-5193}
\affiliation{Ikerbasque, 48011 Bilbao}
%%%\affiliation{University of Illinois at Urbana-Champaign, Urbana, Illinois 61801}
\affiliation{Indian Institute of Technology Guwahati, Assam 781039}
\affiliation{Indian Institute of Technology Madras, Chennai 600036}
%%%\affiliation{Indiana University, Bloomington, Indiana 47408}
\affiliation{Institute of High Energy Physics, Chinese Academy of Sciences, Beijing 100049}
\affiliation{Institute of High Energy Physics, Vienna 1050}
\affiliation{Institute for High Energy Physics, Protvino 142281}
%%%\affiliation{Institute of Mathematical Sciences, Chennai 600113}
%%%\affiliation{INFN - Sezione di Torino, 10125 Torino}
\affiliation{Institute for Theoretical and Experimental Physics, Moscow 117218}
\affiliation{J. Stefan Institute, 1000 Ljubljana}
\affiliation{Kanagawa University, Yokohama 221-8686}
\affiliation{Institut f\"ur Experimentelle Kernphysik, Karlsruher Institut f\"ur Technologie, 76131 Karlsruhe}
%%%\affiliation{Kavli Institute for the Physics and Mathematics of the Universe, University of Tokyo, Kashiwa 277-8583}
\affiliation{Korea Institute of Science and Technology Information, Daejeon 305-806}
\affiliation{Korea University, Seoul 136-713}
%%%\affiliation{Kyoto University, Kyoto 606-8502}
\affiliation{Kyungpook National University, Daegu 702-701}
\affiliation{\'Ecole Polytechnique F\'ed\'erale de Lausanne (EPFL), Lausanne 1015}
\affiliation{Faculty of Mathematics and Physics, University of Ljubljana, 1000 Ljubljana}
\affiliation{Luther College, Decorah, Iowa 52101}
\affiliation{University of Maribor, 2000 Maribor}
\affiliation{Max-Planck-Institut f\"ur Physik, 80805 M\"unchen}
\affiliation{School of Physics, University of Melbourne, Victoria 3010}
\affiliation{Moscow Physical Engineering Institute, Moscow 115409}
\affiliation{Moscow Institute of Physics and Technology, Moscow Region 141700}
\affiliation{Graduate School of Science, Nagoya University, Nagoya 464-8602}
\affiliation{Kobayashi-Maskawa Institute, Nagoya University, Nagoya 464-8602}
%%%\affiliation{Nara University of Education, Nara 630-8528}
\affiliation{Nara Women's University, Nara 630-8506}
\affiliation{National Central University, Chung-li 32054}
\affiliation{National United University, Miao Li 36003}
\affiliation{Department of Physics, National Taiwan University, Taipei 10617}
\affiliation{H. Niewodniczanski Institute of Nuclear Physics, Krakow 31-342}
\affiliation{Nippon Dental University, Niigata 951-8580}
\affiliation{Niigata University, Niigata 950-2181}
%%%\affiliation{University of Nova Gorica, 5000 Nova Gorica}
\affiliation{Osaka City University, Osaka 558-8585}
%%%\affiliation{Osaka University, Osaka 565-0871}
\affiliation{Pacific Northwest National Laboratory, Richland, Washington 99352}
%%%\affiliation{Panjab University, Chandigarh 160014}
\affiliation{Peking University, Beijing 100871}
%%%\affiliation{Punjab Agricultural University, Ludhiana 141004}
\affiliation{Research Center for Electron Photon Science, Tohoku University, Sendai 980-8578}
%%%\affiliation{Research Center for Nuclear Physics, Osaka University, Osaka 567-0047}
%%%\affiliation{RIKEN BNL Research Center, Upton, New York 11973}
%%%\affiliation{Saga University, Saga 840-8502}
\affiliation{University of Science and Technology of China, Hefei 230026}
\affiliation{Seoul National University, Seoul 151-742}
%%%\affiliation{Shinshu University, Nagano 390-8621}
\affiliation{Sungkyunkwan University, Suwon 440-746}
\affiliation{School of Physics, University of Sydney, NSW 2006}
\affiliation{Tata Institute of Fundamental Research, Mumbai 400005}
\affiliation{Excellence Cluster Universe, Technische Universit\"at M\"unchen, 85748 Garching}
\affiliation{Toho University, Funabashi 274-8510}
\affiliation{Tohoku Gakuin University, Tagajo 985-8537}
\affiliation{Tohoku University, Sendai 980-8578}
\affiliation{Department of Physics, University of Tokyo, Tokyo 113-0033}
\affiliation{Tokyo Institute of Technology, Tokyo 152-8550}
\affiliation{Tokyo Metropolitan University, Tokyo 192-0397}
\affiliation{Tokyo University of Agriculture and Technology, Tokyo 184-8588}
%%%\affiliation{Toyama National College of Maritime Technology, Toyama 933-0293}
\affiliation{CNP, Virginia Polytechnic Institute and State University, Blacksburg, Virginia 24061}
\affiliation{Wayne State University, Detroit, Michigan 48202}
\affiliation{Yamagata University, Yamagata 990-8560}
\affiliation{Yonsei University, Seoul 120-749}
  \author{O.~Lutz}\affiliation{Institut f\"ur Experimentelle Kernphysik, Karlsruher Institut f\"ur Technologie, 76131 Karlsruhe} % Karlsruhe
  \author{S.~Neubauer}\affiliation{Institut f\"ur Experimentelle Kernphysik, Karlsruher Institut f\"ur Technologie, 76131 Karlsruhe} % Karlsruhe
  \author{M.~Heck}\affiliation{Institut f\"ur Experimentelle Kernphysik, Karlsruher Institut f\"ur Technologie, 76131 Karlsruhe} % Karlsruhe
  \author{T.~Kuhr}\affiliation{Institut f\"ur Experimentelle Kernphysik, Karlsruher Institut f\"ur Technologie, 76131 Karlsruhe} % Karlsruhe
  \author{A.~Zupanc}\affiliation{Institut f\"ur Experimentelle Kernphysik, Karlsruher Institut f\"ur Technologie, 76131 Karlsruhe} % Karlsruhe
  \author{I.~Adachi}\affiliation{High Energy Accelerator Research Organization (KEK), Tsukuba 305-0801} % KEK
% \author{K.~Adamczyk}\affiliation{H. Niewodniczanski Institute of Nuclear Physics, Krakow 31-342} % Krakow
  \author{H.~Aihara}\affiliation{Department of Physics, University of Tokyo, Tokyo 113-0033} % Tokyo
% \author{K.~Arinstein}\affiliation{Budker Institute of Nuclear Physics SB RAS and Novosibirsk State University, Novosibirsk 630090} % BINP
% \author{Y.~Arita}\affiliation{Graduate School of Science, Nagoya University, Nagoya 464-8602} % Nagoya
  \author{D.~M.~Asner}\affiliation{Pacific Northwest National Laboratory, Richland, Washington 99352} % PNNL
% \author{T.~Aso}\affiliation{Toyama National College of Maritime Technology, Toyama 933-0293} % Toyama
% \author{V.~Aulchenko}\affiliation{Budker Institute of Nuclear Physics SB RAS and Novosibirsk State University, Novosibirsk 630090} % BINP
  \author{T.~Aushev}\affiliation{Institute for Theoretical and Experimental Physics, Moscow 117218} % ITEP
  \author{T.~Aziz}\affiliation{Tata Institute of Fundamental Research, Mumbai 400005} % Tata
% ...
% BINP
% \author{T.~Zivko}\affiliation{J. Stefan Institute, 1000 Ljubljana} % Ljubljana
% \author{N.~Zwahlen}\affiliation{\'Ecole Polytechnique F\'ed\'erale de Lausanne (EPFL), Lausanne 1015} % Lausanne
% \author{O.~Zyukova}\affiliation{Budker Institute of Nuclear Physics SB RAS and Novosibirsk State University, Novosibirsk 630090} % BINP
\collaboration{The Belle Collaboration}

%\input{author}
%\author{Author}\affiliation{affiliation}
%\collaboration{The Belle Collaboration}
%\noaffiliation
%% end author list

\begin{abstract}
We report a search for the rare decays $B\rightarrow h^{(*)}\nu\bar{\nu}$, where $h^{(*)}$ stands for $K^+$, $K_S^0$, $K^{*+}$, $K^{*0}$, $\pi^+$, $\pi^0$, $\rho^+$, $\rho^0$ and $\phi$. 
The results are obtained from a $711$ fb$^{-1}$ data sample 
that contains $772 \times 10^6 B\bar{B}$ pairs collected at the $\Upsilon(4S)$ resonance with the Belle detector at the KEKB $e^+ e^-$
collider. We search for signal candidates by fully reconstructing a hadronic decay of the accompanying $B$ meson and requiring a single $h^{(*)}$ meson left on the signal side. No significant
signal is observed and we set upper limits on the branching fractions at $90\%$ confidence level. The measurements of $B^+\rightarrow K^{*+}\nu\bar{\nu}$, $B^+\rightarrow \pi^+\nu\bar{\nu}$, $B^0\rightarrow \pi^0\nu\bar{\nu}$ and $B^0\rightarrow \rho^0\nu\bar{\nu}$ provide the world's currently most restrictive limits.
\end{abstract}


\maketitle

\tighten

{\renewcommand{\thefootnote}{\fnsymbol{footnote}}}
\setcounter{footnote}{0}

\begin{figure}[htp]
\subfloat[Penguin diagram]{
\centering\includegraphics[width=0.23\textwidth]{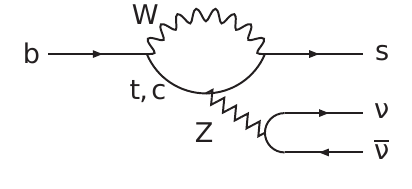}
}
\subfloat[Box diagram]{
\centering\includegraphics[width=0.23\textwidth]{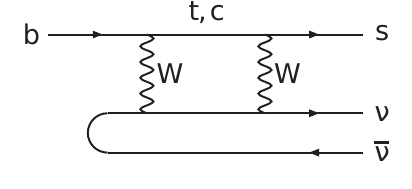}
}
\caption{The quark-level diagrams for the $b\rightarrow s\nu\bar{\nu}$ transition in the standard model.}
\label{feynman}
\end{figure}

The decays $B\rightarrow K^{(*)}\nu\bar{\nu}$ proceed through the flavor-changing neutral-current process $b\rightarrow s\nu\bar{\nu}$, which is sensitive to physics beyond the standard model (SM)~\cite{theoryValues, theorySM}. The dominant SM diagrams are shown in Fig.~\ref{feynman}. The SM branching fractions are estimated to be $(6.8\pm2.0)\times 10^{-6}$ for $B^+\rightarrow K^{*+}\nu\bar{\nu}$~\cite{theorySM} and $(4.4\pm 1.5)\times 10^{-6}$ for $B^+\rightarrow K^{+}\nu\bar{\nu}$ decays~\cite{theoryValues}. The decays $B\rightarrow (\pi,\rho)\nu\bar{\nu}$ proceed similarly through $b\rightarrow d\nu\bar{\nu}$. Compared to $b\rightarrow s\nu\bar{\nu}$ transitions, the branching fractions are further suppressed by a factor $|V_{\rm td}/V_{\rm ts}|^2$. The decay $B^0\rightarrow \phi\nu\bar{\nu}$ proceeds through a yet unobserved penguin annihilation process, with the expected branching fraction thus much lower. The advantage of $\nu\bar{\nu}$ rather than $\ell^+\ell^-$ in the final state is the absence of 
long-distance electromagnetic interactions. In the ratio of the individual branching fractions for $B\rightarrow K\nu\bar{\nu}$ and $B\rightarrow K\ell^+\ell^-$, the form factor normalization cancels out, leading to a factor of three smaller theoretical error compared to the $\nu\bar{\nu}$ mode alone~\cite{theorySM}. Measurements of the $B\rightarrow K\nu\bar{\nu}$ and $B\rightarrow K\ell^+\ell^-$ branching fractions might reveal moderate deviations from SM expectations due to New physics such as SUSY particles, a possible fourth generation and a non-standard $Z$-coupling which would contribute to the penguin loop or box diagram and affect the branching fractions~\cite{theoryNP}.\\

%A theoretical calculation of the decay amplitudes for $B\rightarrow h^{(*)}\nu\bar{\nu}$ is particularly reliable, 
%owing to the absence of long-distance interactions that affect the charged-lepton channels $B\rightarrow h^{(*)}l^+l^-$.

Experimental measurements~\cite{leptons} of the $b\rightarrow s\ell^+\ell^-$ transitions with two charged leptons are in good agreement with SM calculations~\cite{theorySM}. The challenging search for decays with two final-state neutrinos was previously carried out by the CLEO, BaBar and Belle collaborations~\cite{Cleo, Chen, experiment, BaBarNEW}. No signal was observed, and the experimental upper limit for the $B^+\rightarrow K^{+}\nu\bar{\nu}$ decay is a factor of three above the SM prediction; for the other branching fractions, the limits are an order of magnitude above the predictions.\\ 

This measurement of $B\rightarrow h^{(*)}\nu\bar{\nu}$, where $h^{(*)}$ stands for $K^+$, $K_S^0$, $K^{*+}$, $K^{*0}$, $\pi^+$, $\pi^0$, $\rho^+$, $\rho^0$ and $\phi$~\cite{CC}, is based on the full Belle data sample recorded at the $\Upsilon(4S)$ resonance that contains $772 \times 10^6 B\overline{B}$ pairs. The main improvements compared to the previous analysis~\cite{Chen} consist of the use of a new probabilistic full reconstruction, a further optimized background suppression and an improved signal extraction procedure.\\

The Belle detector is a large-solid-angle magnetic spectrometer that consists of a silicon vertex detector (SVD), a 50-layer central drift chamber (CDC), an array of aerogel threshold Cherenkov counters (ACC), a barrel-like arrangement of time-of-flight scintillation counters (TOF), and an electromagnetic calorimeter comprised of CsI(Tl) crystals (ECL) located inside a superconducting solenoid coil that provides a 1.5~T magnetic field. An iron flux return located outside of the coil is instrumented to detect $K_L^0$ mesons and to identify muons (KLM). The detector is described in detail elsewhere~\cite{Belle}. Two inner detector configurations were used. A 2.0 cm radius beam pipe and a 3-layer silicon vertex detector was used for the first sample of $152 \times 10^6 B\bar{B}$ pairs, while a 1.5 cm radius beam pipe, a 4-layer silicon detector and a small-cell inner drift chamber were used to record the remaining $620 \times 10^6 B\bar{B}$ pairs~\cite{svd2}. The data set recorded with the second 
configuration of the SVD was reprocessed with respect to~\cite{Belle} using new track finding algorithms, which improved the track reconstruction efficiency. A GEANT3-based~\cite {geant3} Monte Carlo (MC) simulation of the Belle detector is used to optimize the event selection and to estimate the signal efficiency.\\

We identify signal candidates by fully reconstructing the accompanying $B$ meson ($B_{\rm tag}$) and requiring one single $h^{(*)}$ meson on the signal side. The $B_{\rm tag}$ candidates are reconstructed in hadronic decay channels using a neural network-based hierarchical full reconstruction method~\cite{fullreco}, which provides, for a given purity, roughly twice as many $B_{\rm tag}$ candidates compared to the full reconstruction method used in the previous analysis~\cite{Chen}. The reconstruction is done in four stages; at each stage, the signal probabilities are calculated. In the first stage charged tracks, photons and $K_{s}^0$ and $\pi^0$ mesons are reconstructed. In the following step, two to five of these particles are combined in different modes to form $D^{\pm}_{(s)}$, $D^0$ and $J/\psi$ candidates. Some of the most important variables used in the neural network training are the product of the neural network output for the children, the invariant mass of child 
pairs and the angle between them, the angle between the momentum of the $D$ meson and the vector between the $D$ decay vertex and the interaction point, and the significance of his vector's length. In the third stage, the particles from the prior stages are combined to form the $D^{*\pm}_{(s)}$ and $D^{*0}$ mesons. In the final stage, the $B^{\pm}$ and $B^{0}$ candidates are reconstructed in one of $1104$ exclusive hadronic decay channels. Here, variables with good discrimination power are the product of the neural network outputs of the children, the mass of the $D$ meson, the mass difference of the $D$ and $D^*$ candidates, the angle between the $B$ meson and the thrust axis, and angles between the children. We use the output variable $o_{\rm tag}$ of the full reconstruction, which ranges from zero for background events to unity if a clear $B_{\rm tag}$ is obtained from the network, and require $o_{\rm tag}>0.02$. This cut was found to give the 
best expected branching fraction limit for all channels. We select the $B_{\rm tag}$ candidates using the energy difference $\Delta E \equiv E_{B}-E_{\rm beam}$ and the beam-energy constrained mass $M_{\rm bc} \equiv \sqrt{E_{\rm beam}^2-p_B^2}$, where $E_{\rm beam}$ is the beam energy and $E_B$ and $p_B$ are the reconstructed energy and momentum of $B_{\rm tag}$ 
candidate in the $\Upsilon(4S)$ center-of-mass (c.m.) frame. We require $B_{\rm tag}$ candidates to satisfy the requirements $M_{\rm bc}>5.27$ GeV/$c^2$ and $-0.08$ GeV $<\Delta E<0.06$ GeV. If there are multiple $B_{\rm tag}$ candidates in an event, the candidate with the highest $o_{\rm tag}$ is chosen.\\

The particles in the event not associated with the $B_{\rm tag}$ meson are used to reconstruct a $B_{\rm sig}\rightarrow h^{(*)}\nu\bar{\nu}$ candidate. Prompt charged tracks are required to have a maximum distance to the interaction point (IP) of $5$ cm in the beam direction ($z$), of $2$ cm in the transverse plane ($r-\phi$) and a minimum momentum of $0.1$ GeV/$c$ in the transverse plane. $K^{\pm}$ ($\pi^{\pm}$) candidates are reconstructed from charged tracks having a kaon likelihood greater than $0.6$ (less than $0.4$). The kaon likelihood is defined by $\mathcal{R}_K\equiv\mathcal{L}_K/(\mathcal{L}_K+\mathcal{L_{\pi}})$, where $\mathcal{L}_K$($\mathcal{L_{\pi}}$) denotes a combined likelihood measurement from the ACC, the TOF, and $dE/dx$ from the CDC for the $K^{\pm}$ ($\pi^{\pm}$) tracks. It is a function of the momentum and the polar angle of the tracks in the laboratory frame. The kaon (pion) identification efficiency is $88\%$-$93\%$ ($86\%$-$93\%$) with a pion (kaon) misidentification 
probability of $10\%$-$12\%$ ($8\%$-$11\%$). We use pairs of oppositely charged tracks to reconstruct 
$K_S^0$ decays, with an invariant mass that is within $\pm15$ MeV/$c^2$ of the nominal $K^0_S$ meson mass (corresponding to a width of $5.8\sigma$). We adopt the standard $K^0_S$ selection criteria developed within the Belle collaboration~\cite{kshort}. For $\pi^0\rightarrow\gamma\gamma$, a minimum photon energy of $50$ MeV is required and the $\gamma\gamma$ invariant mass must be within $\pm16$ MeV/$c^2$ of the nominal $\pi^0$ mass ($4.1\sigma$).\\

The decays $B_{\rm sig}^+\rightarrow K^+\nu\bar{\nu}$, $B_{\rm sig}^+\rightarrow \pi^+\nu\bar{\nu}$, $B_{\rm sig}^0\rightarrow K^0_S\nu\bar{\nu}$ and $B_{\rm sig}^0\rightarrow \pi^0\nu\bar{\nu}$ are reconstructed from single $K^+$, $\pi^+$, $K_S^0$ and $\pi^0$ candidates, respectively. The $B_{\rm sig}^0\rightarrow K^{*0}\nu\bar{\nu}$ candidates are reconstructed from a charged pion and an oppositely charged kaon, while $B_{\rm sig}^+\rightarrow K^{*+}\nu\bar{\nu}$ decays are reconstructed from a $K_S^0$ candidate and a charged pion, or a $\pi^0$ candidate and a charged kaon. The reconstructed mass of the $K^{*0}$($K^{*+}$) candidate is required to be within a $\pm75$ MeV/$c^2$ window around the nominal $K^{*0}$($K^{*+}$) mass. Furthermore, pairs of charged pions with opposite charge are used to form $B_{\rm sig}^0\rightarrow \rho^0\nu\bar{\nu}$ candidates, where the $\pi^+\pi^-$ invariant mass must be within $\pm150$ MeV/$c^2$ of the nominal $\rho^0$ mass. For $B_{\rm sig}^+\rightarrow \rho^+\nu\bar{\nu}$, 
a charged pion and a $\pi^0$ candidate within a $\pm150$ MeV/$c^2$ mass window around the nominal $\rho^+$ mass are used. A $K^+K^-$ pair with a reconstructed mass within $\pm10$ MeV/$c^2$ of the nominal $\phi$ mass is used to reconstruct $\phi$ candidates. After identifying the $B_{\rm tag}$ candidate and reconstructing the light meson, we require that no additional charged tracks nor $\pi^0$ candidates remain in the event. These vetoes and the mutually exclusive PID requirements for kaons and pions also eliminate the possibility of obtaining multiple $h^{(*)}$ candidates per event.\\

The dominant backgrounds are from $e^+e^-\rightarrow q \bar{q}$ ($q=u,d,s,c$) continuum events and $B\bar{B}$ decays with a $b\rightarrow c$ transition. During the full reconstruction, a continuum suppression algorithm based on modified Fox-Wolfram moments~\cite{KSFW} is applied. To further suppress the continuum background, we use the cosine of the angle between the momentum of the $h^{(*)}$ and the thrust axis with the sign convention to the side of
momentum flow of the rest of the charged tracks, evaluated in the $\Upsilon(4S)$ rest frame. This cosine is close to $-1$ or $1$ for continuum events but uniformly distributed for spherical $B\bar{B}$ events. We require the cosine to lie between $-0.8$ and $0.7$. The selection criteria are asymetric due to the kinematic selection performed during the $B_{\rm tag}$ reconstruction. In this way, the continuum background component is nearly completely removed from the signal region, which leads to a better signal sensitivity compared to the previous analysis~\cite{Chen}.\\
~\\
We introduce a lower bound of $1.6$ GeV/$c$ on the momentum of the $h^{(*)}$ candidate in the $B_{\rm sig}$ rest frame to suppress the background from $b\rightarrow c$ transitions. An upper bound of $2.5$ GeV/$c$ rejects the contributions from radiative two-body modes such as $B\rightarrow K^*\gamma$. The momentum requirement is removed for $\phi$ candidates due to the lack of theoretical calculations for $B \rightarrow \phi$ form factors. To suppress backgrounds with undetected particles produced along the beam pipe, we require the cosine of the angle between the missing momentum in the laboratory frame and the beam to lie between $-0.86$ and $0.95$. Contributions from rare $B$ decays involving $b\rightarrow u$, $b\rightarrow s$, or $b\rightarrow d$ processes are found to be small according to MC studies. The only exception is the $B_{\rm sig} \rightarrow \phi \nu\bar{\nu}$ decay, where rare decays represent the majority of the remaining background events. The $B^+\rightarrow\tau^+\nu_{\tau}$ decay with 
the $\pi^+ \nu \bar{\nu}$ and $\rho^+ \nu \bar{\nu}$ final states contributes only $3\%$ and $2\%$ of the total background in these channels, respectively.\\
~\\
The efficiency of the full reconstruction differs between data and MC simulation. The correction ratio, depending on the $B_{\rm tag}$ decay mode and obtained from a study using $b \to c$
semileptonic decays on the signal side, lies between $0.7$ and $0.8$ and is applied to all correctly reconstructed $B_{\rm tag}$ candidates in the MC simulation.\\
~\\
The most powerful variable to identify the signal decays is the residual energy in the ECL, $E_{\rm ECL}$, which is the sum of the energies of ECL clusters that are not associated with the $B_{\rm tag}$ daughters nor with the signal-side $h^{(*)}$ candidate. To suppress contributions from noise in the calorimeter, minimum energy thresholds are required: $50$ MeV for the barrel, $100$ MeV for the forward endcap and $150$ MeV for the backward endcap region. These thresholds were determined and optimized to achieve an optimal signal to noise ratio in the calorimeter crystals. In a properly reconstructed signal event, no activity should appear in the calorimeter, so signal events peak at low $E_{\rm ECL}$ values.\\
\begin{figure}[htp]
\subfloat{
\centering\includegraphics[width=0.22\textwidth]{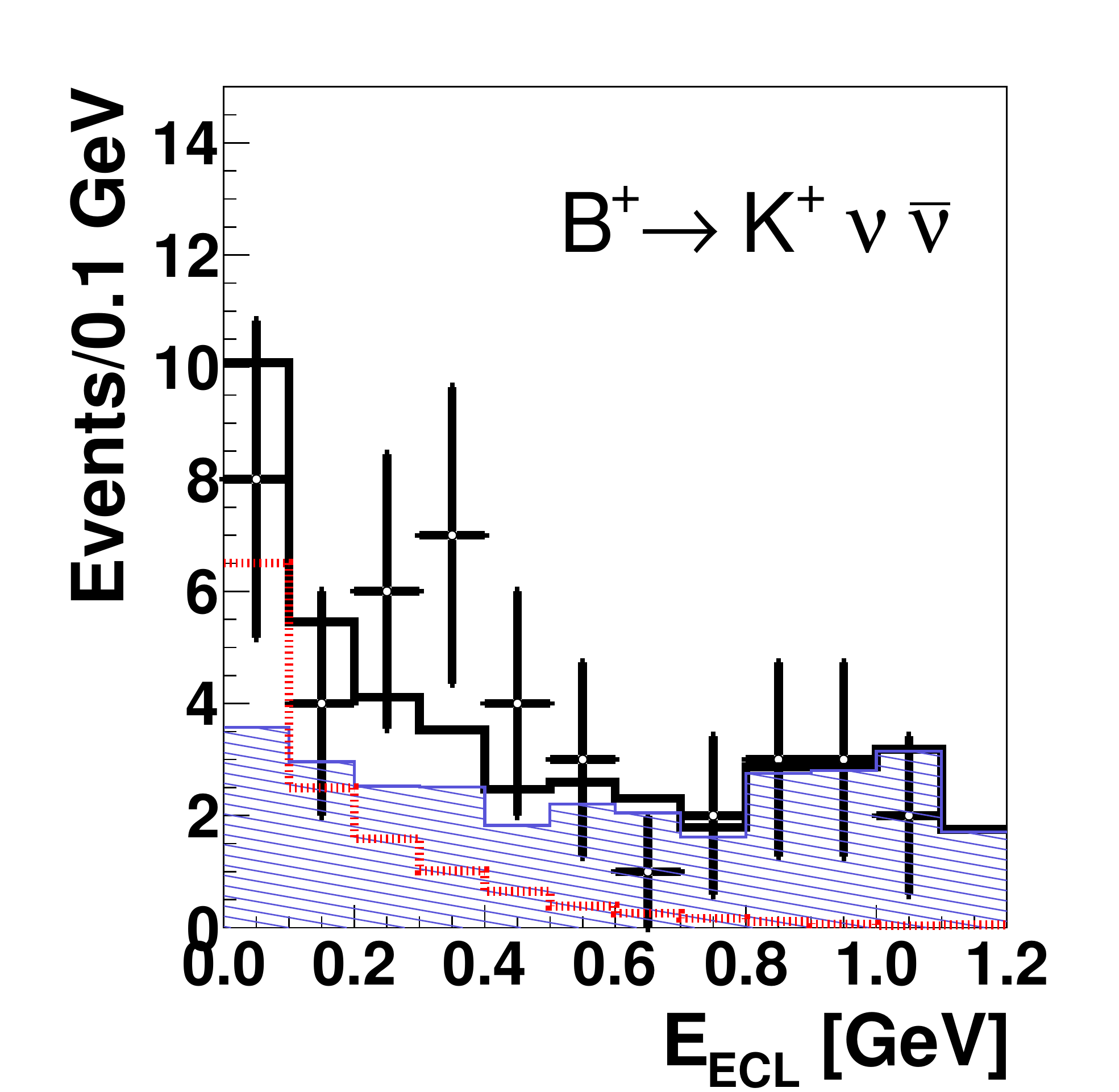}
}
\subfloat{
\centering\includegraphics[width=0.22\textwidth]{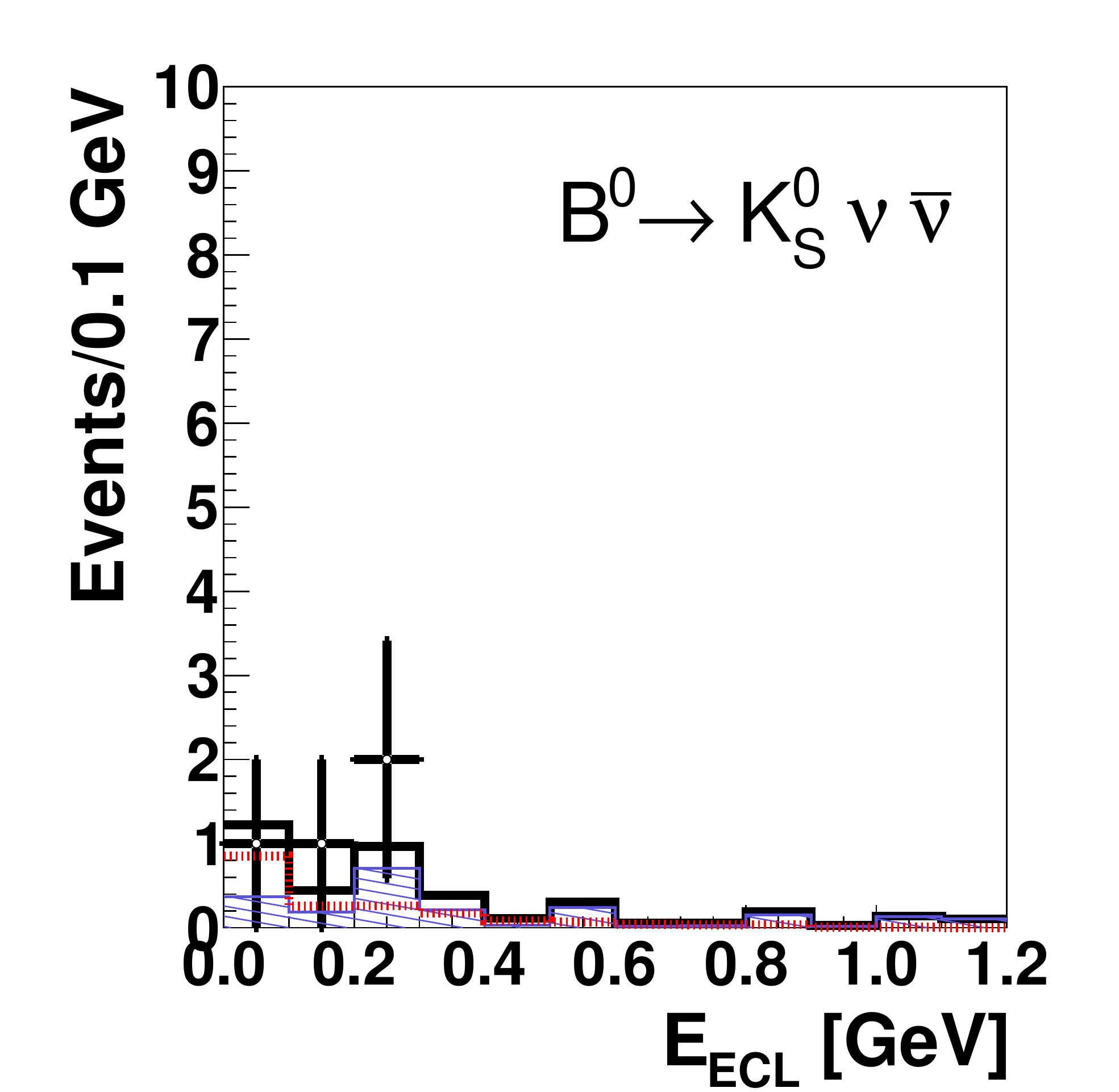}
}

\subfloat{
\centering\includegraphics[width=0.22\textwidth]{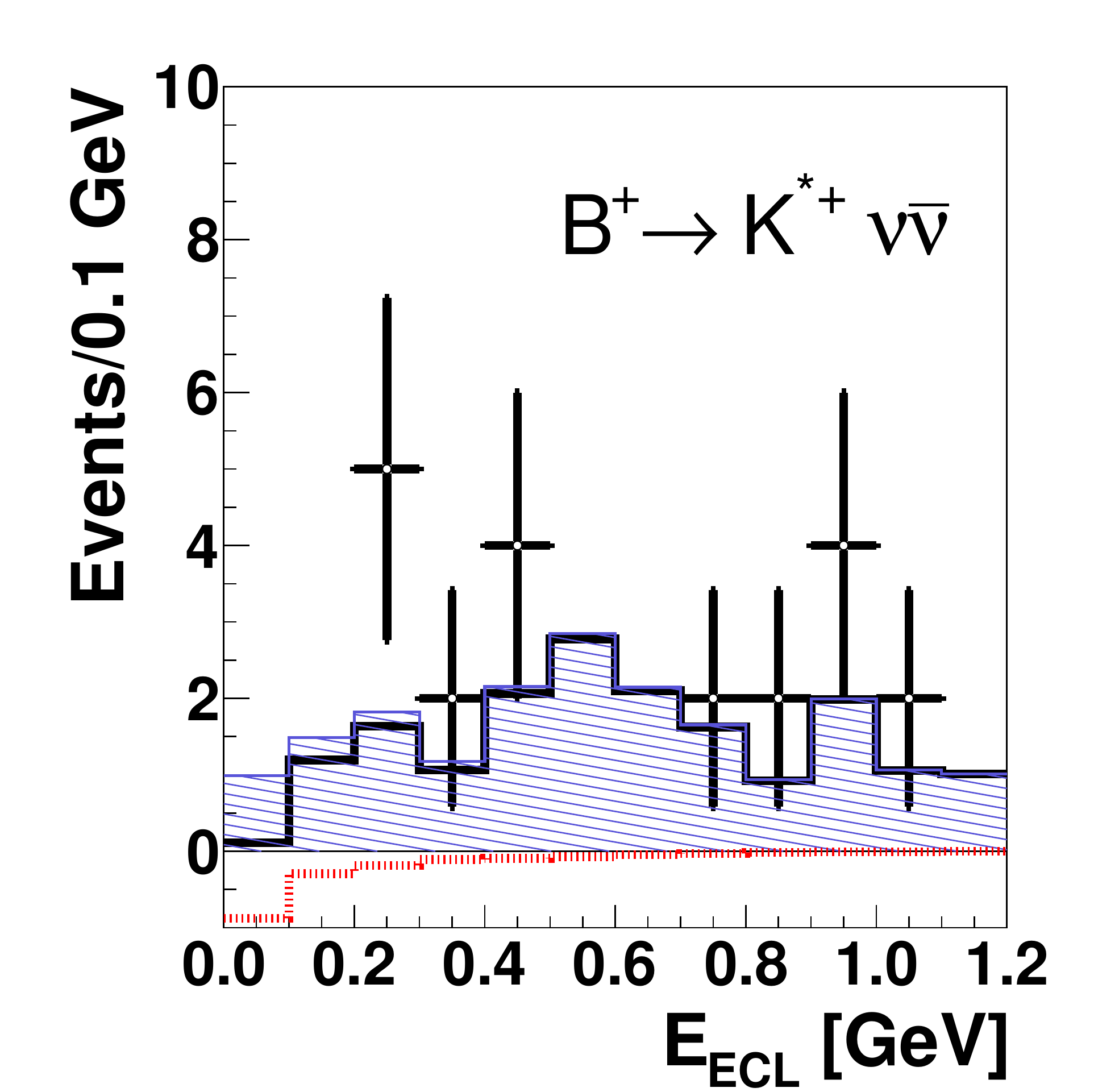}
}
\subfloat{
\centering\includegraphics[width=0.22\textwidth]{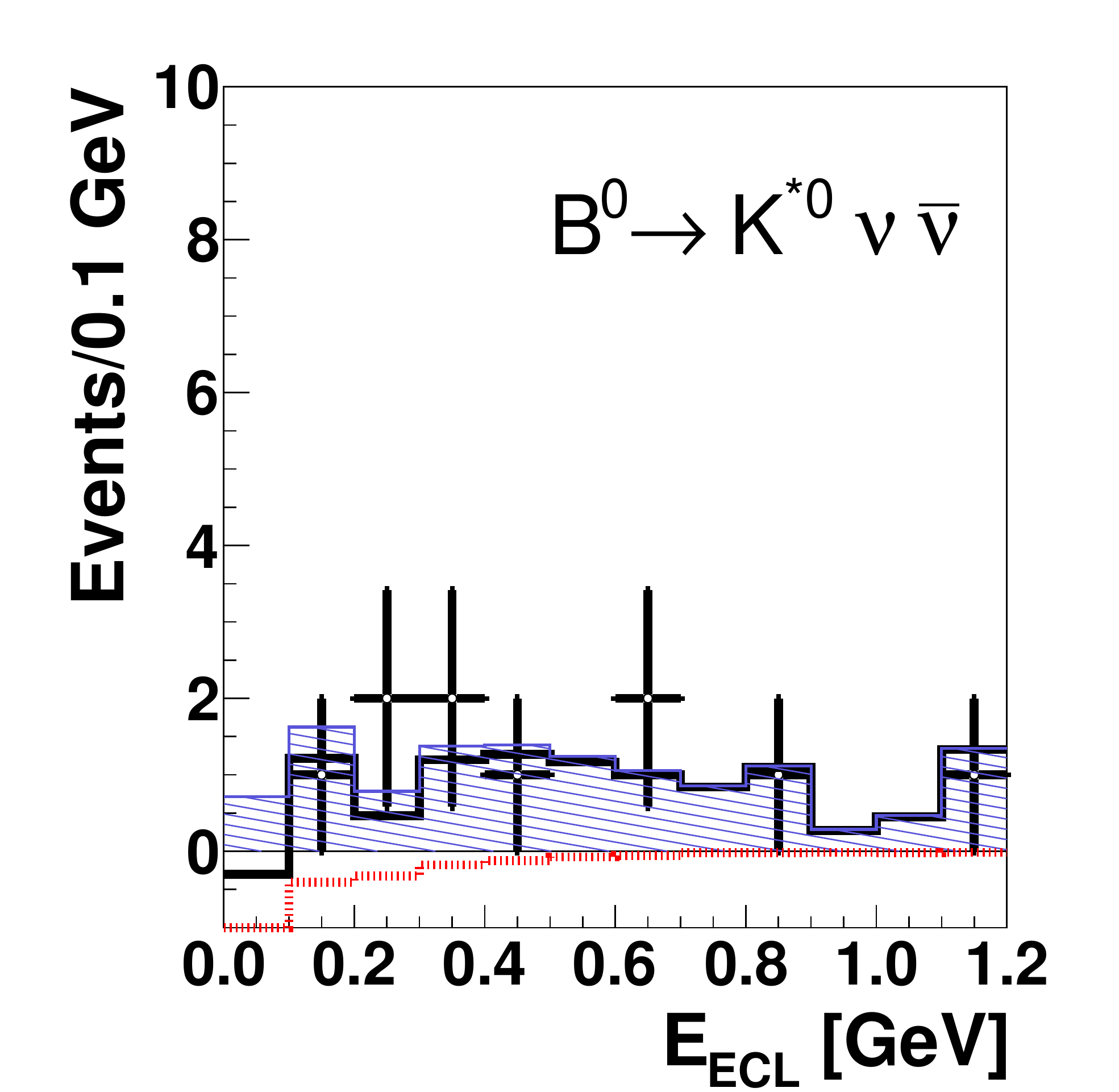}
}

\subfloat{
\centering\includegraphics[width=0.22\textwidth]{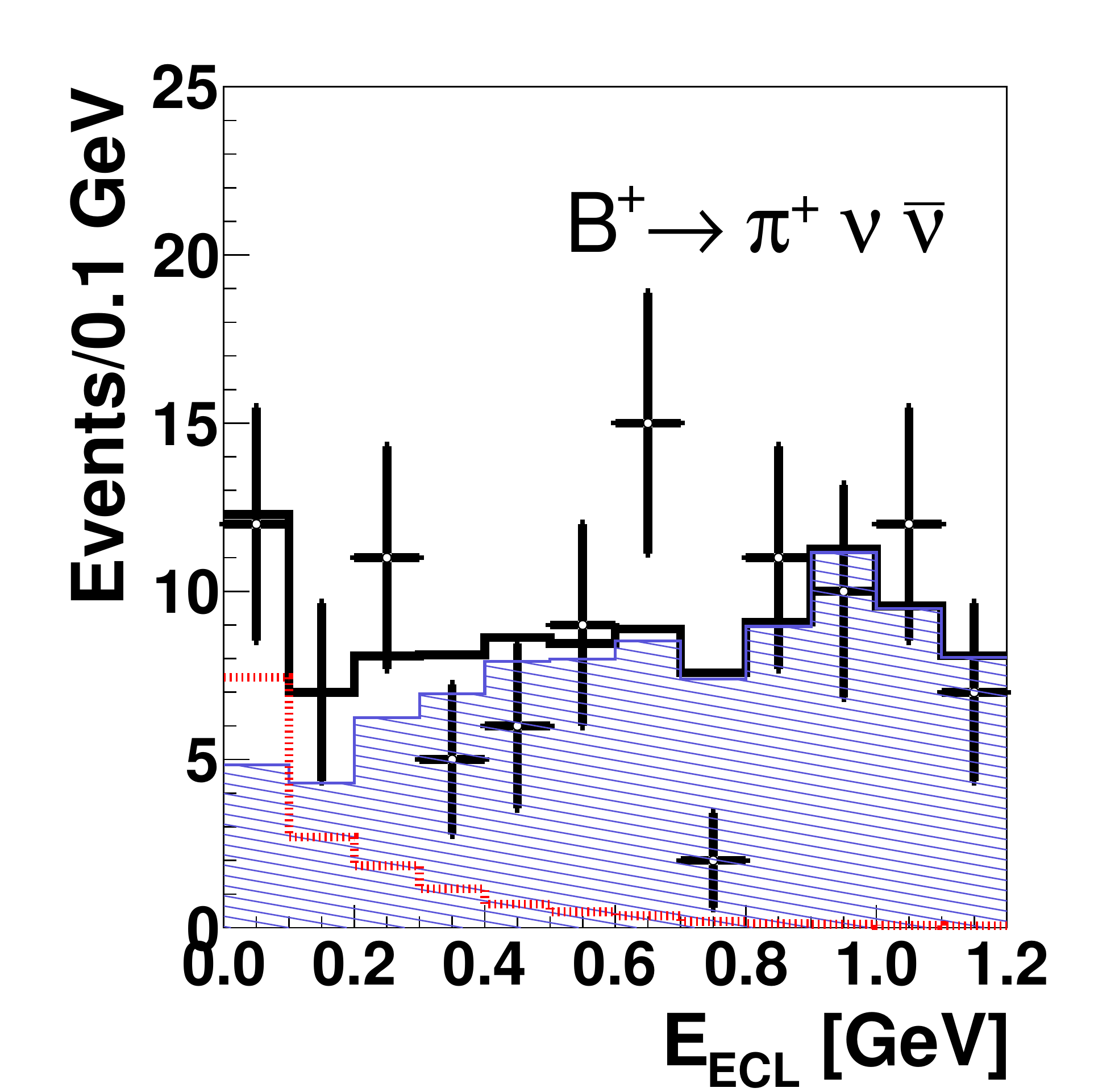}
}
\subfloat{
\centering\includegraphics[width=0.22\textwidth]{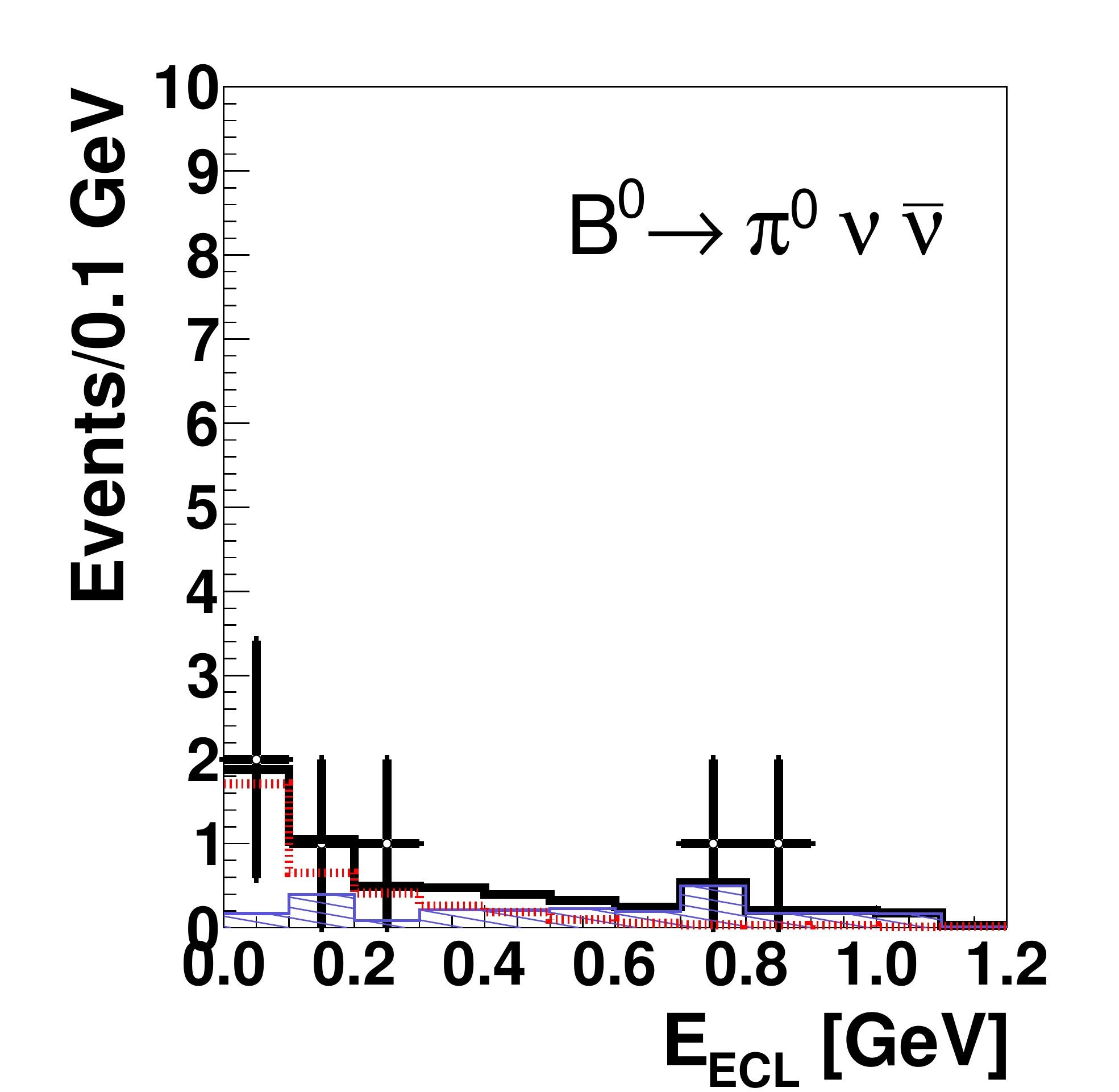}
}

\subfloat{
\centering\includegraphics[width=0.22\textwidth]{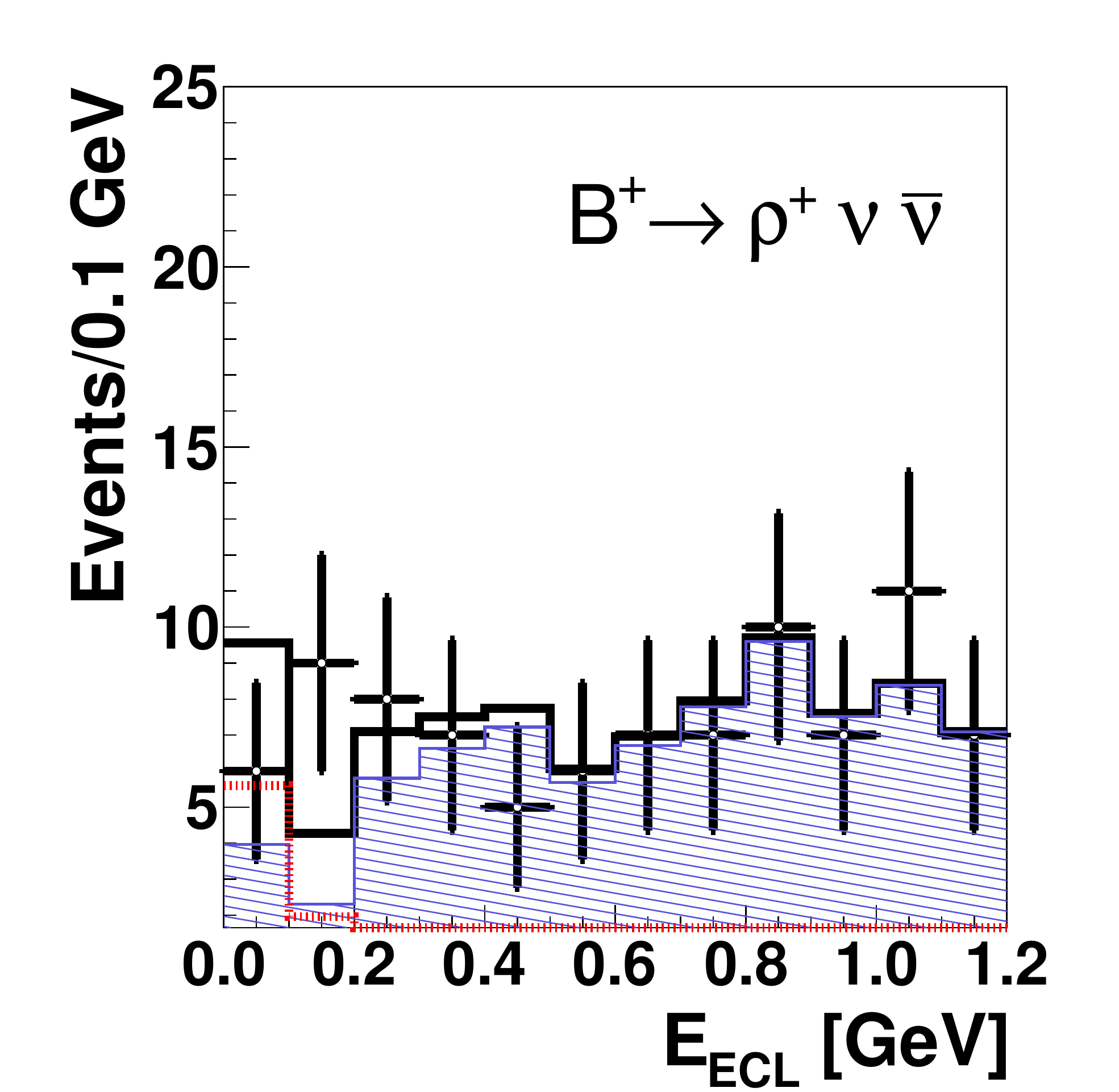}
}
\subfloat{
\centering\includegraphics[width=0.22\textwidth]{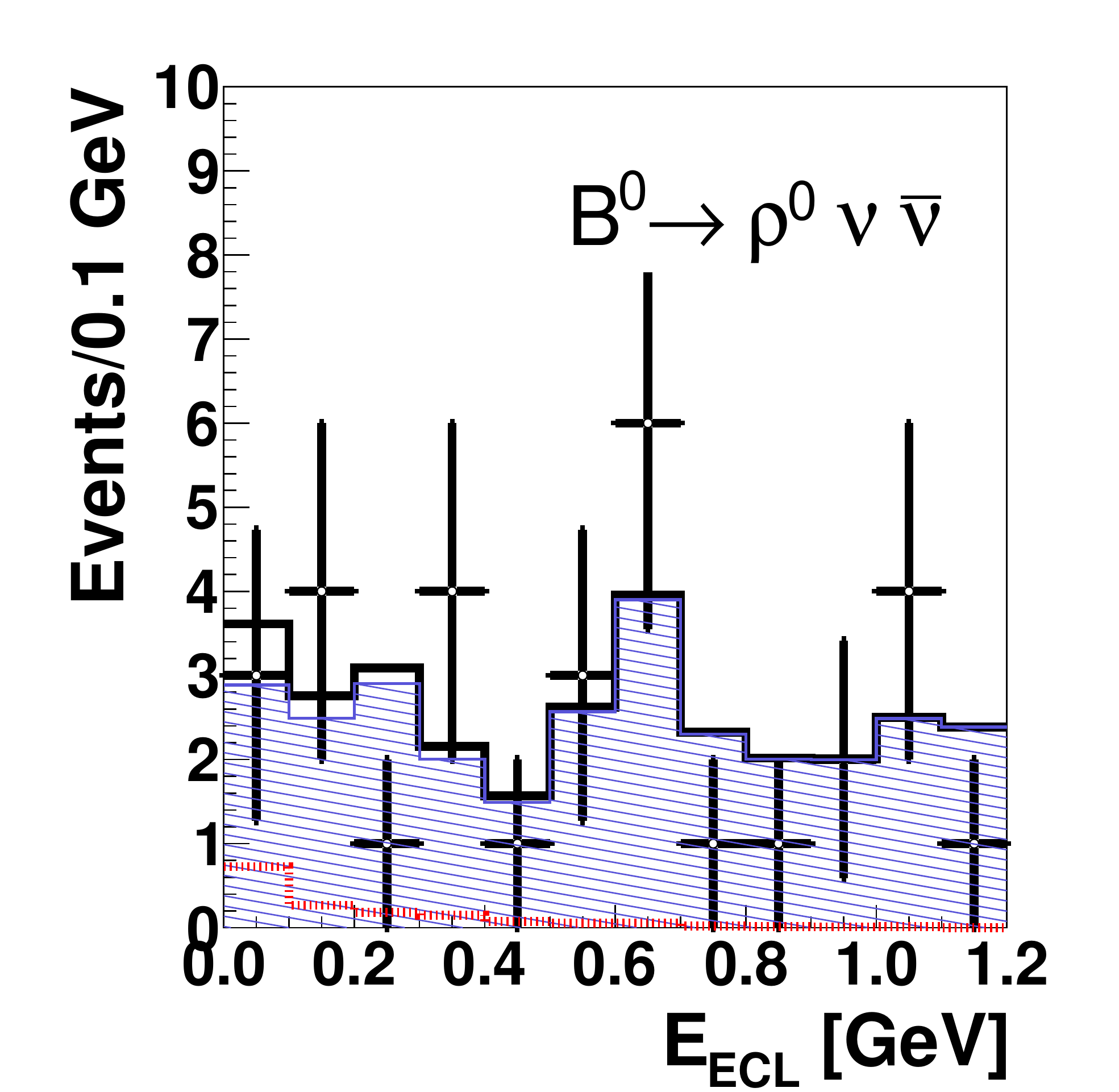}
}

\subfloat{
\centering\includegraphics[width=0.22\textwidth]{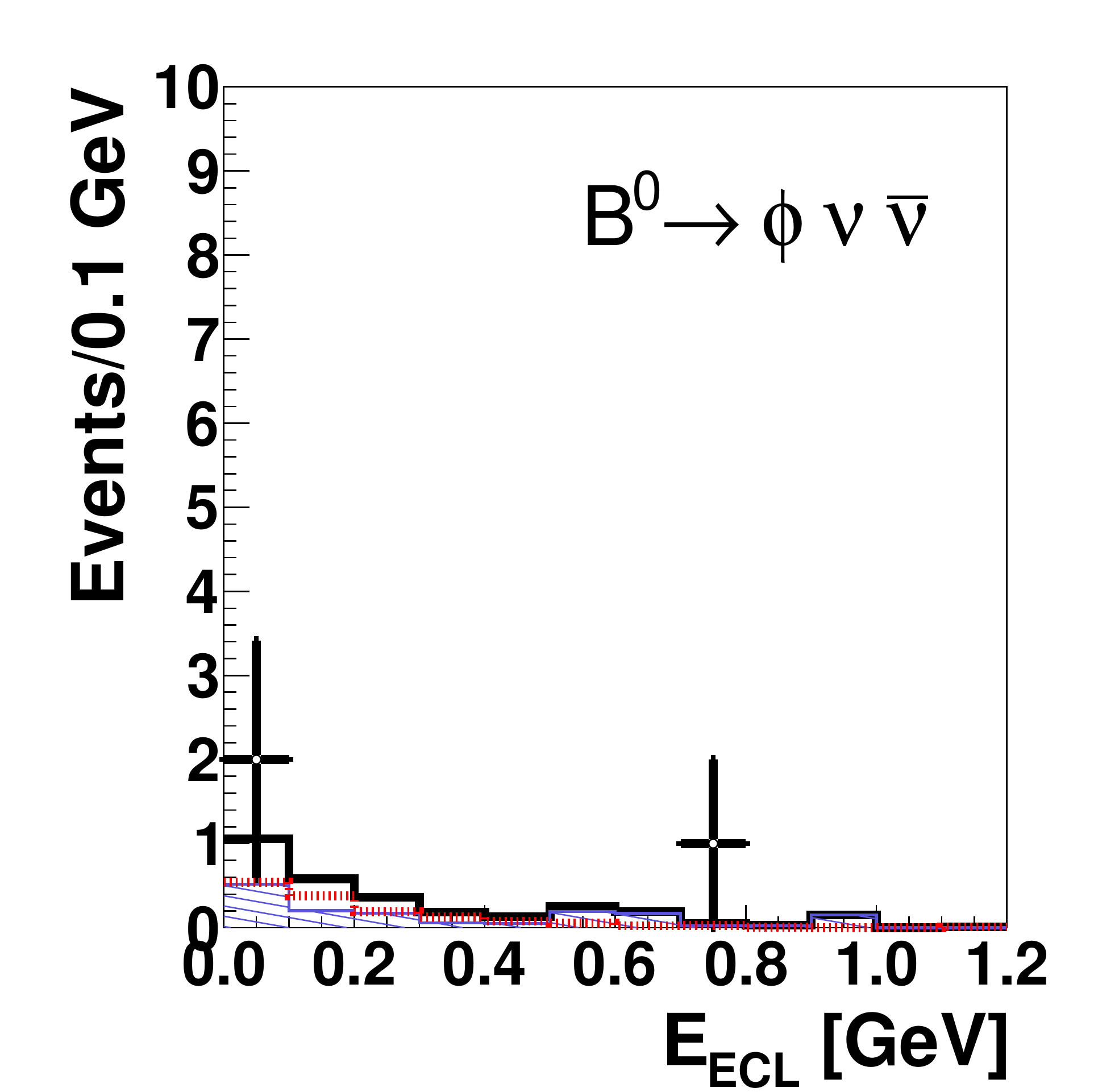}
}
\caption{The $E_{\rm ECL}$ distributions for $B\rightarrow h^{(*)}\nu\bar{\nu}$ decays. Points with error bars are data; the solid black histogram is the total fit result. The blue cross-hatched region is the background component; the dashed red histogram shows the signal contribution.}
\label{resultsPlot}
\end{figure}

The signal yield is extracted from an extended binned maximum likelihood fit to the $E_{\rm ECL}$ distribution in the range from $0$ to $1.2$ GeV. The likelihood is
\begin{equation}
  {\cal L}=\frac{(\sum_j N_j)^{N}\quad e^{-\sum_j N_j}}{N!}\quad\prod_{i=1}^N\sum_j N_j{\cal P}_j^i \quad ,
\end{equation}
where $N$ is the total number of observed events, $N_{j}$ is the yield for category $j$, which corresponds to either signal or background, $i$ is the event index and $\mathcal{P}_j$ is the probability density function (PDF) of the fit component $j$. The PDFs are obtained using MC simulation and are modeled as histogram functions. The normalizations of single background components (continuum, $b\rightarrow c$, and rare) relative to each other are estimated from the MC simulation and fixed in the fit, leading to two free parameters in the fit for signal and total background yields. Compared to the counting procedure performed in the previous analysis~\cite{Chen}, this fitting method makes use of the signal shape to discriminate between signal and background contributions.\\
~\\
We calculate the significances and the upper limits by evaluating the likelihood profile. To take into account the systematic uncertainty, we convolve the likelihood function with a Gaussian whose width equals the total systematic uncertainty. The significance is obtained by comparing the likelihood values at maximum and at zero signal yield: $S = \sqrt{2\log\left(\frac{\mathcal{L}_{\rm max}}{\mathcal{L}_0}\right)}$. The upper limit on the branching fraction at $90\%$ confidence level (C.L.) is evaluated through a Bayesian method by integrating the likelihood function from zero to the bound that gives $90\%$ of the total area; this assumes a uniform prior distribution for the branching fraction. We obtain the branching fraction using the signal yield $N_{\rm sig}$, the signal efficiency $\epsilon$ and the total number of $B\bar{B}$ pairs $N_{B\bar{B}}$: $\mathcal{B} = N_{\rm sig} /( \epsilon \cdot N_{B\bar{B}})$.\\
~\\
To evaluate the sensitivity, simulated experiments with the expected amount of background events and zero signal events were generated. For each of the experiments, an upper limit on the branching fraction at $90\%$ C.L. was calculated. The median values of the obtained upper limit distributions are summarized in the rightmost column in Table~\ref{fitResults}.\\

\renewcommand{\arraystretch}{1.5}
\begin{table*}[htb]
\caption{Summary of the total number of observed events in the signal box $N_{\rm tot}$, the resulting signal yield $N_{\rm sig}$, the significance of the observed signal, the signal efficiencies $\epsilon$, the measured and the expected upper limits on the branching fractions at $90\%$ C.L.}
\label{fitResults}
\begin{tabular}
 {@{\hspace{0.3cm}}l@{\hspace{0.3cm}}  @{\hspace{0.3cm}}c@{\hspace{0.3cm}} @{\hspace{0.3cm}}c@{\hspace{0.3cm}} @{\hspace{0.3cm}}c@{\hspace{0.3cm}} @{\hspace{0.3cm}}c@{\hspace{0.3cm}} @{\hspace{0.3cm}}c@{\hspace{0.4cm}}|@{\hspace{0.4cm}}c@{\hspace{0.3cm}}}
\hline \hline
Mode&$N_{\rm tot}$ &$N_{\rm sig}$& Significance& $\epsilon, 10^{-4}$ &Upper limit& Expected limit\\
\hline
$B^+\rightarrow K^{+}\nu\bar{\nu}$&$43$&$13.3^{+7.4}_{-6.6}(\rm stat)\pm2.3(\rm syst)$&$2.0\sigma$&$5.68$&$<5.5\times10^{-5}$&$2.2\times10^{-5}$\\
$B^0\rightarrow K_s^{0}\nu\bar{\nu}$&$4$&$1.8^{+3.3}_{-2.4}(\rm stat)\pm1.0(\rm syst)$&$0.7\sigma$&$0.84$&$<9.7\times 10^{-5}$&$7.3\times 10^{-5}$\\
$B^+\rightarrow K^{*+}\nu\bar{\nu}$&$21$&$-1.7^{+1.7}_{-1.1}(\rm stat)\pm1.5(\rm syst)$&--&$1.47$&$<4.0\times10^{-5}$&$5.8\times10^{-5}$\\
$B^0\rightarrow K^{*0}\nu\bar{\nu}$&$10$&$-2.3^{+10.2}_{-3.5}(\rm stat)\pm 0.9(\rm syst)$&--&$1.44$&$<5.5\times10^{-5}$&$4.6\times10^{-5}$\\
$B^+\rightarrow \pi^{+}\nu\bar{\nu}$&$107$&$15.2^{+7.1}_{-6.2}(\rm stat)\pm1.4(\rm syst)$&$2.6\sigma$&$3.39$&$<9.8\times 10^{-5}$&$3.8\times 10^{-5}$\\
$B^0\rightarrow \pi^{0}\nu\bar{\nu}$&$6$&$3.5^{+2.6}_{-1.9}(\rm stat)\pm0.6(\rm syst)$&$1.9\sigma$&$1.66$&$<6.9\times10^{-5}$&$3.6\times10^{-5}$\\
$B^+\rightarrow \rho^{+}\nu\bar{\nu}$&$90$&$11.3^{+6.3}_{-5.4}(\rm stat)\pm4.1(\rm syst)$&$1.7\sigma$&$1.35$&$<21.3\times 10^{-5}$&$10.2\times 10^{-5}$\\
$B^0\rightarrow \rho^{0}\nu\bar{\nu}$&$31$&$1.6^{+5.0}_{-4.1}(\rm stat)\pm0.4(\rm syst)$&$0.4\sigma$&$0.64$&$<20.8\times10^{-5}$&$15.7\times10^{-5}$\\
$B^0\rightarrow \phi\nu\bar{\nu}$&$3$&$1.4^{+2.9}_{-0.9}(\rm stat)\pm0.8(\rm syst)$&$0.5\sigma$&$0.58$&$<12.7\times10^{-5}$&$8.7\times10^{-5}$\\
\hline \hline
\end{tabular}
\end{table*}
\renewcommand{\arraystretch}{1.0}

\renewcommand{\arraystretch}{1.5}
\begin{table*}[htb]
\caption{Summary of the systematic errors. The errors on the signal yield are given in the number of events and the errors of the signal normalization are given in $\%$.}
\label{systematicsSum}
\begin{tabular} {@{\hspace{0.25cm}}l@{\hspace{0.25cm}}  @{\hspace{0.25cm}}c@{\hspace{0.25cm}} @{\hspace{0.25cm}}c@{\hspace{0.25cm}} @{\hspace{0.25cm}}c@{\hspace{0.25cm}} @{\hspace{0.25cm}}c@{\hspace{0.25cm}} @{\hspace{0.25cm}}c@{\hspace{0.25cm}} @{\hspace{0.25cm}}c@{\hspace{0.25cm}}  @{\hspace{0.25cm}}c@{\hspace{0.25cm}} @{\hspace{0.25cm}}c@{\hspace{0.25cm}} @{\hspace{0.25cm}}c@{\hspace{0.25cm}} }
\hline \hline
\small{Channel}&$K^{+}\nu\bar{\nu}$&$K_s^{0}\nu\bar{\nu}$&$K^{*+}\nu\bar{\nu}$&$K^{*0}\nu\bar{\nu}$&$\pi^{+}\nu\bar{\nu}$&$\pi^{0}\nu\bar{\nu}$&$\rho^{+}\nu\bar{\nu}$&$\rho^{0}\nu\bar{\nu}$&$\phi\nu\bar{\nu}$\\
\hline
\it Signal yield [events]&&&&&&&&&\\
\hline
%Background model&$15.7$&$53.0$&$89.0$&$24.0$&$6.1$&$10.6$&$35.4$&$22.7$&$32.8$\\
%Fit bias&--&--&$7.9$&$28$&--&$11.7$&--&$7.3$&$44.0$\\
Background model&$2.1$&$0.9$&$1.5$&$0.5$&$0.9$&$0.4$&$4.0$&$0.4$&$0.5$\\
Fit bias&--&--&$0.2$&$0.6$&--&$0.4$&--&$0.1$&$0.6$\\
%\hline
%Sum&$2.1$&$0.9$&$1.5$&$0.8$&0.5&$0.6$&$4.0$&$0.4$&$0.8$\\
\hline
\hline
\it Signal normalization [$\%$]&&&&&&&&&\\
\hline
Track and $\pi^0$ rejection&$4.4$&$4.4$&$4.4$&$4.4$&$4.4$&$4.4$&$4.4$&$4.4$&$4.4$\\
$B_{\rm tag}$ correction&$4.2$&$4.5$&$4.2$&$4.5$&$4.2$&$4.5$&$4.2$&$4.5$&$4.5$\\
Signal MC statistics&$1.2$&$3.5$&$3.7$&$2.8$&$1.5$&$2.1$&$2.3$&$3.3$&$2.6$\\
Track, $\pi^0$ and $K^0_S$ reconstruction efficiency&$0.3$&$2.3$&$4.1$&$0.4$&$0.4$&$4.0$&$4.2$&$0.7$&$1.4$\\
Particle identification&$2.0$&$4.0$&$2.0$&$4.0$&$2.0$&--&$2.0$&$4.0$&$4.0$\\
$N_{B\bar{B}}$&$1.4$&$1.4$&$1.4$&$1.4$&$1.4$&$1.4$&$1.4$&$1.4$&$1.4$\\
Form factors&$2.0$&$5.4$&$3.8$&$6.4$&$1.9$&$1.6$&$2.9$&$4.5$&$7.5$\\
%\hline
%Sum&$9.7$&$11.2$&$11.1$&$10.7$&$9.8$&$10.6$&$10.8$&$10.9$&$10.8$\\
\hline
 \hline
\end{tabular}
\end{table*}
\renewcommand{\arraystretch}{1.0}

The $E_{\rm ECL}$ distributions in data are shown in Fig.~\ref{resultsPlot}, superimposed with the fit result. The total numbers of observed events, the signal yields, the significances of the observed signal, the reconstruction efficiencies and the upper limits on the branching fractions are summarized in Table~\ref{fitResults}. None of the signal modes show a significant signal contribution. According to MC studies, the enhancements in the $K^+\nu\bar{\nu}$ and $\pi\nu\bar{\nu}$ modes 
are unlikely to be caused by peaking background contributions. The signal reconstruction efficiencies are estimated with MC simulations using the $B\rightarrow h^{(*)}$ form factors from Ref.~\cite{formfactors}.\\

The systematic uncertainty is dominated by the statistical uncertainty of the background model. The stringent selection increases the signal to background ratio but also reduces the number of MC events in the signal box. This leads to a large uncertainty in the background shape, despite using an MC sample corresponding to five times the data luminosity. To estimate the uncertainty, we replace the nominal background model with two alternative models compatible with the simulation and repeat the fit. The alternative background models are Chebyshev polynomials of order $0$, $1$ or $2$. For each channel, the two models that are most compatible with the background distribution are used. After the fit with these models, the largest deviation of the signal yield from the nominal fit is assigned as systematic error, which can vary in size among channels due to the different background shapes. To validate the procedure we also performed a crosscheck for one of the channels by refitting the sample with randomly fluctuating background histogram models and obtained a compatible result. The fit bias is evaluated through pseudo-experiments with signal and background yields set to the observed values. The 
systematic uncertainty due to MC data discrepancy of the track and $\pi^0$ rejection was studied using a $D^{(*)}l\nu$ control sample. Uncertainties associated with the $B_{\rm tag}$ reconstruction efficiency, signal MC statistics, particle identification, track or particle reconstruction efficiency, the total number of the $B\bar{B}$ pairs and the form factors of the signal model are included as well. Because the limits quoted in this paper are derived under the 
assumption a SM signal distribution, the signal model uncertainty is obtained from the uncertainties of the SM prediction of the form factors. As the main
impact of the different form factors relates to the $h^{(*)}$ momentum distribution, the evaluated systematic error also includes the impact of a different $h^{(*)}$ momentum distribution. 
The systematic effects of the $o_{\rm tag}$ and $\Delta E$ cuts, as well as the minimum energy thresholds in the calorimeter were found to be negligible. All systematic uncertainties are summarized in Table~\ref{systematicsSum}. The total systematic uncertainty is calculated by summing all contributions in quadrature and is generally smaller than the statistical error.\\
~\\
In conclusion, we have performed a search for $B\rightarrow h^{(*)}\nu\bar{\nu}$ decays in nine different modes with a fully reconstructed $B_{\rm tag}$ on a data sample of $772 \times 10^6 B\bar{B}$ pairs collected at the $\Upsilon(4S)$ resonance with the Belle detector. No significant signal is observed and we set upper limits on the branching fraction at $90\%$ C.L. The results of this analysis supersede the previous results from Belle~\cite{Chen}. The limits reported here for $K^{*+}\nu\bar{\nu}$, $\pi^+\nu\bar{\nu}$, $\pi^0\nu\bar{\nu}$ and $\rho^0\nu\bar{\nu}$ are the most stringent constraints to date~\cite{PDG}. These limits are above SM predictions and allow room for new physics contributions. The upcoming Belle II experiment~\cite{belle2} should be able to reach a sensitivity high enough to probe the SM predictions for exclusive $b\rightarrow s \nu\bar{\nu}$ decays.\\

We thank the KEKB group for excellent operation of the
accelerator; the KEK cryogenics group for efficient solenoid
operations; and the KEK computer group, the NII, and 
PNNL/EMSL for valuable computing and SINET4 network support.  
We acknowledge support from MEXT, JSPS and Nagoya's TLPRC (Japan);
ARC and DIISR (Australia); NSFC (China); MSMT (Czechia); the Carl Zeiss Foundation and the DFG (Germany);
DST (India); INFN (Italy); MEST, NRF, GSDC of KISTI, and WCU (Korea); 
MNiSW (Poland); MES and RFAAE (Russia); ARRS (Slovenia); 
SNSF (Switzerland); NSC and MOE (Taiwan); and DOE and NSF (USA).


\begin{thebibliography}{99}
\bibitem{CC}
Throughout this paper, 
the inclusion of the charge-conjugate decay mode is implied
unless otherwise stated.
\end{thebibliography}
\end{document}